\documentclass[10pt,twocolumn,letterpaper,twocolumn]{article}
\usepackage{ol2}
\usepackage[draft]{hyperref}
\usepackage{amsmath}
\usepackage{calc}
\usepackage{amsfonts}
\usepackage{amssymb}
\usepackage{graphicx}
\usepackage{bm}
\usepackage{color}
\usepackage{epsfig}
\usepackage{ifthen}

\begin{document}

\twocolumn[
\title{Nonlinear adiabatic optical isolator}

\author{Andon Rangelov$^{1,*}$ and Stefano Longhi$^{2}$}

\address{
$^1$Department of Physics, Sofia University, James Bourchier 5 blvd., 1164 Sofia, Bulgaria\\
$^2$Dipartimento di Fisica, Politecnico di Milano and Istituto di Fotonica e Nanotecnologie del Consiglio Nazionale delle Ricerche, Piazza L. da Vinci 32, I-20133 Milano, Italy\\
$^*$Corresponding author: rangelov@phys.uni-sofia.bg}

\begin{abstract}
We theoretically propose a method for optical isolation based on
adiabatic nonlinear sum frequency generation in a chirped
quasi-phase-matching crystal  with strong absorption at the
generated sum frequency wave. The method does not suffer from
limitations of dynamic reciprocity found in other nonlinear optical
isolation methods, and can provide tunable optical isolation with
ultrafast all-optical switching capability. Moreover, as an
adiabatic technique it is robust to variations in the optical design
and is relatively broadband.\end{abstract}


%(190.0190) Nonlinear optics;
%(190.4223) Nonlinear wave mixing;
%(190.2620) Harmonic generation and mixing;
%(260.1180) Crystal optics
% 190.3100   Instabilities and chaos
% 270.3430   Laser theory
% 140.3560   Lasers, ring
% 140.5960   Semiconductor lasers

 ] %% activate for two-column option

An optical isolator (optical diode) is the optical correspondent of
electronic diode, allowing unidirectional non-reciprocal light
transmission. These devices are widely used in optical
telecommunications and laser applications to prevent the unwanted
feedback that might be harmful to optical instruments and devices.
Moreover, the use of an isolator generally improves the performance
of an optical circuit as it suppresses spurious interferences,
interactions between different devices and undesired light routing
\cite{jalas2013}. Currently, optical diodes rely almost exclusively
on the Faraday effect where external magnetic fields are used to
break time reversal symmetry \cite{Aplet1964}. However, optical
isolators based on the Faraday effect are typically large-size
devices,  they cannot be implemented easily in on-chip integrated
systems \cite{Bi}, and do not provide dynamic optical isolation with
all-optical switching capability, which is desirable in advanced
optical signal processing.

Several recent works have suggested and experimentally demonstrated
new ways
to realize optical isolators that do not rely on magneto-optical effects \cite%
{Trevino1996,Gallo2001,Soljacic,Hwang2005,Yu2009,Doerr2011,Doerr2013,Kang2011,Dong2016,Lepri2011,Fan2012,Fan2013,Bender2013,Nazari2014,Peng2014,Chang2014}.
Dynamic modulation methods, which enable tunable optical isolation
with all-optical switching capabilities, have been recently proposed
and experimentally demonstrated using liquid-crystal heterojunctions
\cite{Hwang2005}, phase modulators in InP and silicon photonics
\cite{Doerr2011,Doerr2013}, opto-acoustic photonic crystal fibers
\cite{Kang2011}, and traveling-wave Mach-Zehnder modulators
\cite{Dong2016}. A broad class of optical isolators is the one based
on nonlinear interaction of light in a nonlinear medium, which
breaks Lorentz reciprocity. Nonlinear optical
isolators have raised great attention since more than two decades \cite%
{Trevino1996,Gallo2001,Soljacic,Lepri2011,Fan2012,Fan2013,Bender2013,Nazari2014,Peng2014,Chang2014}%
, mainly because of their all-optical switching capability
exploiting the ultrafast response of the nonlinear medium and for
on-chip integration possibilities. However, their effectiveness in
optical isolation has been recently questioned owing of the
appearance of so-called dynamical reciprocity \cite{Shi2015}. In a
nonlinear optical isolator, non-reciprocal transmission contrast is
observed when strong waves are injected in forward or backward
directions, however optical isolation is constrained by a
reciprocity relation for a class of small-amplitude additional waves
that can be transmitted.\par In this Letter we suggest a different
route toward nonlinear optical isolation, which does not suffer from
dynamical reciprocity limitations. The method is based on three wave
mixing process in a second-order nonlinear medium with a strong
control pump wave at frequency $\omega_1$, a weak signal wave at
frequency $\omega_2$, and large absorption at the generated
frequency $\omega_3= \omega_1 \pm \omega_2$. The undepleted pump
approximation linearizes the three wave mixing process and thus our
nonlinear optical diode does not suffer from dynamical reciprocity.
Furthermore since the nonlinear response of the medium is very fast
(instantaneous), switching on and off the pump wave results in
on/off diode action, allowing for ultrafast all-optical switching
isolation capability. Finally to make the optical diode robust and
relatively broadband we use adiabatic frequency conversion in
aperiodically-poled quasi-phase-matched (QPM) crystals, a technique
recently demonstrated by
Suchowski et al. \cite%
{Suchowski2008,Suchowski2009,Suchowski2011,Moses,Suchowski2013}. The
feasibility of the method is illustrated for sum frequency
generation in potassium titanyl phosphate (KTP) crystals, showing
the possibility of achieving an optical isolation up to $\sim 40$ dB
over more than 50 nm in the near-infrared.

%***************************************************************
\begin{figure}[tbh]
\centerline{\includegraphics[width=1\columnwidth]{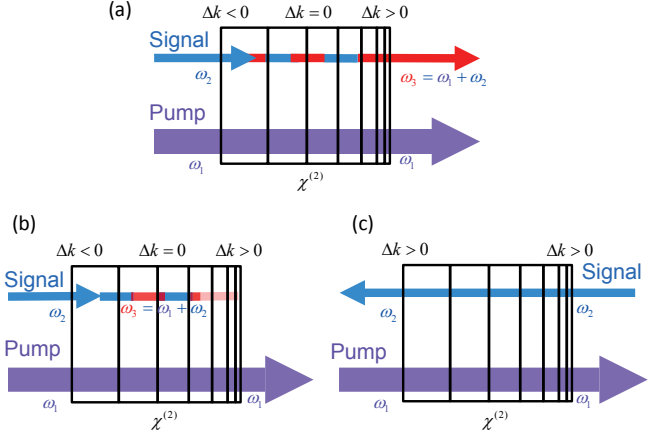}}
\caption{(Color online) Principle of a nonlinear adiabatic optical
diode. Adiabatic SFG schemes with no absorption at the generated
frequency (a) and with absorption at the generated frequency (b) and
(c). (a) Slowly changing the poling period along the crystal length
ensures broadband and robust generation of the SFG wave. (b) For a
forward propagating signal wave, the broadband SFG wave is absorbed
in the crystal, and no transmission occurs. (c) Reversing the
propagation direction of the signal breaks the symmetry and as a
result there is no interaction of the contra propagating pump and
signal waves. Therefore (b) together with (c) work as a broadband
optical diode. } \label{Fig1}
\end{figure}
%***************************************************************
The starting point of our analysis is provided by a standard model
of sum frequency generation (SFG) or frequency difference generation
(DFG) in a nonlinear second-order crystal with a chirped QPM
grating, in which the SFG (or DFG) wave experiences strong linear
absorption during propagation along the crystal. For the sake of
definiteness, we will consider here the SFG case, however a similar
analysis holds for the DFG scheme. In the undepleted pump
approximation, SFG is described by the linear coupled equations \cite%
{Suchowski2014,Boyd}
\begin{subequations}
\label{Sum-Frequency Generation}
\begin{eqnarray}
i\frac{d}{dx}\tilde{A}_{2} &=&\frac{1}{2}\Omega
_{2}\tilde{A}_{3}\exp \left(
-i\int_{0}^{x}\Delta k(\xi )d\xi \right) , \\
i\frac{d}{dx}\tilde{A}_{3} &=&\frac{1}{2}\Omega
_{3}\tilde{A}_{2}\exp \left( i\int_{0}^{x}\Delta k(\xi )d\xi \right)
-i\frac{\Gamma }{2}\tilde{A}_{3} \;\;\;
\end{eqnarray}%
\end{subequations}
where $\tilde{A}_{2}$, $\tilde{A}_{3}$ are the slowly-varying
amplitudes of signal and SFG waves at frequencies $\omega _{2}$ and
$\omega _{3}=\omega _{1}+\omega _{2}$, respectively, $\omega _{1}$
is the frequency of the pump beam with undepleted amplitude
$\tilde{A}_{1}$, $\Delta k(x)$ is the effective residual local phase
mismatch that accounts for the chirped QPM
grating, ${\Omega }_{m}=\chi ^{\left( 2\right) }\omega _{m}\tilde{A}%
_{1}/(2cn_{\omega _{m}})$ are the coupling coefficients, $n_{\omega _{l}}$ ($%
l=1,2,3$) is the refractive indices of the crystal at frequency
$\omega _{l}$ ($l=1,2,3$), $\chi ^{\left( 2\right) }$ is the
effective second order susceptibility of the crystal, $x$ is the
position along the propagation axis, $c$ is the speed of light in
vacuum, and $\Gamma $ is the absorption
coefficient of the SFG wave. After the substitution $\tilde{A}_{2}=A_{2}%
\sqrt{\Omega _{2}}\exp \left[ -i\int_{0}^{x}d\xi \Delta k(\xi )/2\right] $, $%
\tilde{A}_{3}=A_{3}\sqrt{\Omega _{3}}\exp \left[ i\int_{0}^{x}d\xi
\Delta k(\xi )/2\right] $ Eq.(1) can be cast in the form
\begin{equation}
i\frac{d}{dx}\left(
\begin{array}{c}
A_{2} \\
A_{3}%
\end{array}%
\right) =\frac{1}{2}\left(
\begin{array}{cc}
-\Delta k & \Omega \\
\Omega & \Delta k-i\Gamma /2%
\end{array}%
\right) \left(
\begin{array}{c}
A_{2} \\
A_{3}%
\end{array}%
\right) ,  \label{Sum-Frequency Generation 2}
\end{equation}%
where $\Omega =\sqrt{\Omega _{2}\Omega _{3}}$. Interestingly, Eq. (\ref%
{Sum-Frequency Generation 2}) can be viewed as a photonic analogue
of a two-level atomic system which interacts with an external
chirped pulsed
field, with the excited state decaying out of the system with a decay rate $%
\Gamma $ \cite{Akulin92,Vitanov97,Shore2006}. For a linear chirp
$\Delta k(x)=\alpha x$, Eq. (\ref{Sum-Frequency Generation 2})
describes the dissipative Landau-Zener model which admits of an
exact solution in terms of parabolic cylinder functions
\cite{Vitanov97}. The non-vanishing absorption $\Gamma $ at the SFG
wave basically annihilates the signal wave while being converted in
the nonlinear crystal, thus preventing forward propagation, while
the chirped QPM grating ensures
broadband and robust optical isolation under adiabatic operation. The limiting case $%
\Gamma =0$, previously considered in Refs.
\cite{Suchowski2008,Suchowski2009,Suchowski2014},
 realizes broadband SFG via adiabatic rapid passage under the
adiabatic condition
\begin{equation}
\left\vert \Omega \frac{d}{dx}\Delta k\right\vert \ll \left( \Omega
^{2}+\Delta k^{2}\right) ^{3/2}.  \label{adiabatic condition}
\end{equation}%
Such a condition requires a smooth $x$ variation of the phase
mismatch $\Delta k(x)$, i.e. a sufficiently small gradient $\alpha
$, and large coupling $\Omega $. In this way broadband, robust and
almost $\sim 100\%$ conversion efficiency
of the injected signal wave into the SFG wave can be obtained; see Fig.\ref%
{Fig1}(a). To realize an optical diode, a relatively strong
absorption at the generated SFG wave should be considered.
Absorption together with the pump field direction break mirror
symmetry and the optical transmission of the signal wave at
frequency $\omega _{2}$ becomes non reciprocal, as schematically
shown in Figs.\ref{Fig1}(b) and (c). In fact, in the forward
propagation direction phase matching among signal, pump and SFG wave
is realized, the signal wave is converted into the SFG via rapid
adiabatic passage and the SFG is fully absorbed [Fig.\ref{Fig1}(b)].
On the other hand, for backward propagation of the signal wave the
phase matching for frequency conversion is not realized and the
crystal turns out to fully transmit the signal wave
[Fig.\ref{Fig1}(c)]. To realize the nonlinear optical diode, some
constraints should be met for the absorption coefficient, crystal
length and adiabatic rate of the the QPM grating in order to satisfy
the adiabatic condition  (3) and to avoid the so-called overdamping
problem, i.e. phase mismatch induced by too strong absorption. Such
constraints were investigated in details  in the study of the
dissipative Landau-Zener model \cite{Akulin92,Vitanov97}. In
addition to the adiabatic condition (\ref{adiabatic condition}) the
absorption coefficient should be strong enough at the generated SFG
wave in order to suppress it, but too strong absorption coefficient
will have opposite effect as pointed by Vitanov and Stenholm
\cite{Vitanov97}: a too strong absorption freezes the dynamics like
in an overdamped oscillator. One can view the effect of the large
absorption rate as similar to that of large phase mismatching: both
effectively decouple the interaction in three wave mixing. The
optimum regime for the optical diode is realized when the absorption
rate is of the same order as the coupling, i.e. $\Gamma \approx
\Omega $ \cite{Vitanov97}.
%***************************************************************
\begin{figure}[tbh]
\centerline{\includegraphics[width=1\columnwidth]{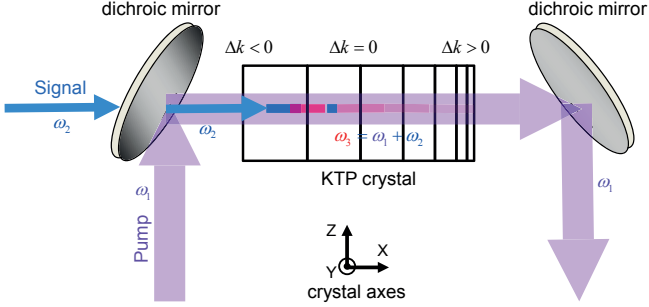}}
\caption{(Color online) Schematic of the nonlinear adiabatic optical
diode realized in a KTP waveguide with a linearly-chirped QPM
grating. The strong pump wave at frequency $\protect\omega _{1}$ and
the weak signal wave at frequency $\protect\omega _{2}$ are
polarized in $z$ direction and phase matched in the KTP crystal to
generate the SFG wave at frequency $\protect\omega
_{3}=\protect\omega _{1}+\protect\omega _{2}$, which is strongly
absorbed by the crystal.} \label{Fig2}
\end{figure}
%***************************************************************
%***************************************************************
\begin{figure}[tbh]
\centerline{\includegraphics[width=0.9\columnwidth]{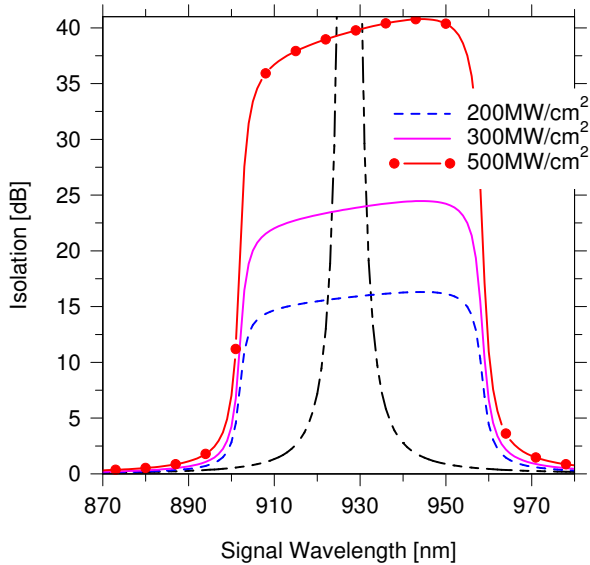}}
\caption{(Color online) Numerically-computed isolation spectra of
the nonlinear adiabatic isolator for three different pump
intensities $I_1=200, 300$ and 500 MW/cm$^2$ versus wavelength of
the signal wave for a 5-cm-long KTP crystal. The isolation of a
5-cm-long phase-matched crystal at 930 nm signal with constant
poling period of 1.79 $\protect\mu $m is plotted by black
dashed-dotted curve for easy reference.} \label{Fig3}
\end{figure}
%***************************************************************

\begin{figure}[tbh]
\centerline{\includegraphics[width=0.9\columnwidth]{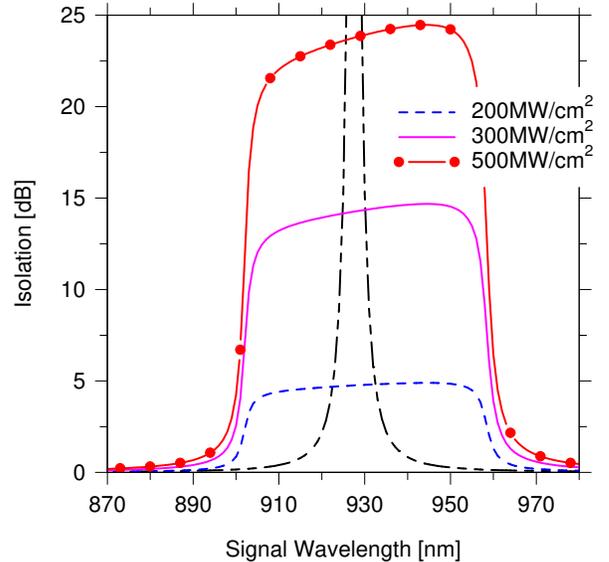}}
\caption{(Color online) Same ad Fig.3, but for a 3-cm-long KTP
crystal.} \label{Fig4}
\end{figure}

To illustrate the feasibility of the proposed method and to provide
design parameter of optimized optical isolation based on the
phootnic analogue of the dissipative Landau-Zener model, let us
consider SFG in potassium titanyl phosphate KTiOPO$_{4}$ (KTP)
crystal. KTP crystals are commonly used in nonlinear optics
applications, showing high damage threshold, a high nonlinear
optical coefficient  and strong absorption in the near ultraviolet
spectral range \cite{Nikogosyan,Kato}. In waveguide configuration,
relatively long crystals can be manufactured with small mode area to
ensure high intensity and diffraction-free long interaction lengths
\cite{uff1,uff2}. We consider specifically type-0 SFG at room
temperature, with a strong pump wave at wavelength $\lambda_1=532$
nm and a weak signal wave at around $\lambda_2=930$ nm,
corresponding to a SFG wave at $\lambda_3=338$ nm. The optical waves
propagate along  the $x$ optical axis and all electric fields are
polarized in $z$ direction\ of the crystal (Fig.\ref{Fig2}), while
phase matching over a bandwidth of more than $ 50$ nm is achieved
using a linearly-chirped QPM grating. The absorption coefficient of
KTP at the SFG wave $\lambda_3 \sim 340$ nm is $\Gamma \sim 229$
cm$^{-1}$ \cite{Dudelzak}. In our simulations, we assume a strong
pump wave at $\lambda_1=532$ nm with an intensity $I_1$ up to
300-500 MW/cm$^{2}$, which provides high efficiency frequency
conversion avoiding the overdamping problem discussed above. For a
KTP waveguide with an effective mode area $A_1 \sim 10 \; \mu$m$^2$,
a pump intensity of $300$ MW/cm$^2$ corresponds to an optical pump
power $P_1 =A_1 I_1 \sim 30$ W, which is available both in
continuous-wave or pulsed regimes from frequency-doubled high-power
Nd:YAG lasers \cite{Mukhopadhyay}. A crystal length of 3-5 cm is
typically assumed, with poling period of the chirped grating varying
from 1.7 $\mu $m to 1.9 $\mu $m along the sample for first-order
QPM. Such grating periods are feasible with current poling
techniques in KTP \cite{Zukauskas}. The performance of the nonlinear
optical diode is provided by the isolation parameter, defined as
\begin{equation}
dB=10  \times {\rm Log}_{10}\left( \frac{T_{f}}{T_{b}}\right) ,
\end{equation}%
where $T_{f}$ and $T_{b}$ are the transmitted electric field
intensities in the forward and backward directions, respectively.
Figures \ref{Fig3} and \ref{Fig4} show the numerically-computed
isolation parameter of the adiabatic nonlinear diode
 for 5-cm and 3-cm-long KTP crystals, respectively. As can be seen from
the figure, a maximum isolation of $\sim 40$ dB and good isolation
$>$35dB over a spectral region of 60 can be obtained in the longer
crystal configuration.

%***************************************************************
%***************************************************************

In conclusion, we have presented a novel route toward the
realization of broadband and tunable nonlinear optical isolators
with ultrafast all-optical switching capabilities, that do not
suffer from dynamical reciprocity commonly found in nonlinear
optical isolation schemes \cite{Shi2015}. Our method is based on
adiabatic frequency conversion (SFG) in aperiodically-poled
quasi-phase-matched crystals with strong absorption at the generated
SFG wave. Ultrafast tunability of optical isolation can be simply
achieved by changing the intensity level of the strong pump wave.
Frequency conversion in the chirped nonlinear crystal is described
by an effective dissipative Landau-Zener model
\cite{Akulin92,Vitanov97,Shore2006}, efficient frequency conversion
and absorption of the SFG wave requiring a balance between
absorption coefficient and nonlinear coupling. The feasibility of
the method has been discussed by considering as an example optical
isolation in the near-infrared ($ \sim 900$ nm)
using a KTP crystal with a chirped QPM grating pumped at 532 nm (frequency-doubled Nd:YAG laser). Optical isolation up to $\sim 40$ dB over more that 50 nm bandwidth has been obtained in numerical simulations. Such a relatively strong and broadband optical isolation indicates that nonlinear optical schemes can provide viable routes toward tunable optical isolation with ultrafast switching capabilities, without being limited by dynamical reciprocity \cite{Shi2015}.\\
\\
We acknowledge support by program DRILA.

\end{document}